\documentclass[twocolumn,prl,floatfix,citeautoscript,nofootinbib,superscriptaddress]{revtex4}
\usepackage{amsbsy}
\usepackage{latexsym,epsfig,graphicx}
\usepackage{dcolumn}
\usepackage{graphicx}
\usepackage{subfigure}
\usepackage{comment}
\usepackage{color}
\usepackage{bm}
\usepackage{mathrsfs}
\usepackage{amsfonts}
\usepackage{amsmath}
\usepackage{color}
\usepackage{amssymb}
\usepackage{xspace}
\usepackage{epstopdf}
\usepackage{tabularx}
\usepackage{longtable}
\usepackage[colorlinks=true, letterpaper=true, pdfstartview=FitV, linkcolor=blue, citecolor=blue, urlcolor=blue]{hyperref}
\usepackage[normalem]{ulem}

\pdfoutput=1
\input{tcilatex}
\begin{document}

\title{In-Plane Zeeman Field-Induced Majorana Corner and Hinge Modes in an $%
s $-Wave Superconductor Heterostructure}
\author{Ya-Jie Wu}
\affiliation{Department of Physics, The University of Texas at Dallas, Richardson, Texas
75080-3021, USA}
\affiliation{School of Science, Xi'an Technological University, Xi'an 710032, China}
\author{Junpeng Hou}
\affiliation{Department of Physics, The University of Texas at Dallas, Richardson, Texas
75080-3021, USA}
\author{Yun-Mei Li}
\affiliation{Department of Physics, The University of Texas at Dallas, Richardson, Texas
75080-3021, USA}
\author{Xi-Wang Luo}
\thanks{xiwang.luo@utdallas.edu}
\affiliation{Department of Physics, The University of Texas at Dallas, Richardson, Texas
75080-3021, USA}
\author{Xiaoyan Shi}
\affiliation{Department of Physics, The University of Texas at Dallas, Richardson, Texas
75080-3021, USA}
\author{Chuanwei Zhang}
\thanks{chuanwei.zhang@utdallas.edu}
\affiliation{Department of Physics, The University of Texas at Dallas, Richardson, Texas
75080-3021, USA}

\begin{abstract}
Second-order topological superconductors host Majorana corner and hinge
modes in contrast to conventional edge and surface modes in two and three
dimensions. However, the realization of such second-order corner modes
usually demands unconventional superconducting pairing or complicated
junctions or layered structures. Here we show that Majorana corner modes
could be realized using a 2D quantum spin Hall insulator in proximity
contact with an $s$-wave superconductor and subject to an in-plane Zeeman
field. Beyond a critical value, the in-plane Zeeman field induces opposite
effective Dirac masses between adjacent boundaries, leading to one Majorana
mode at each corner. A similar paradigm also applies to 3D topological
insulators with the emergence of Majorana hinge states. Avoiding complex
superconductor pairing and material structure, our scheme provides an
experimentally realistic platform for implementing Majorana corner and hinge
states.
\end{abstract}

\maketitle

{\color{blue}\emph{Introduction.{---}}} Majorana zero energy modes in
topological superconductors and superfluids \cite%
{Hansan2010,Qix2011,Elliott2015,Ramon2017} have attracted great interest in
the past two decades because of their non-Abelian exchange statistics and
potential applications in topological quantum computation \cite%
{Kitaev2003,Nayak2008}. A range of physical platforms \cite%
{Oppen2010,Lutchyn2010,Alicea2011,Jiang2011,Alicea2012,Tewari2012,Fu2008,zhang2008,Sato2009,Buhler2014,Ortiz2018,Varona2018}
in both solid state and ultracold atomic systems have been proposed to host
Majorana modes. In particular, remarkably experimental progress has been
made recently to observe Majorana zero energy modes in \textit{s}-wave
superconductors in proximity contact with materials with strong spin-orbit
coupling, such as semiconductor thin films and nanowires, topological
insulators, etc. \cite%
{Mourik2012,Finck2013,Nadj2014,Xuj2015,Wang2016,Heq2017}. In such
topological superconductors and superfluids, Majorana zero energy modes
usually localize at 2D vortex cores or 1D edges, where the Dirac mass in the
low-energy Hamiltonian changes sign.

\begin{figure}[t]
\centering\includegraphics[width=0.35\textwidth]{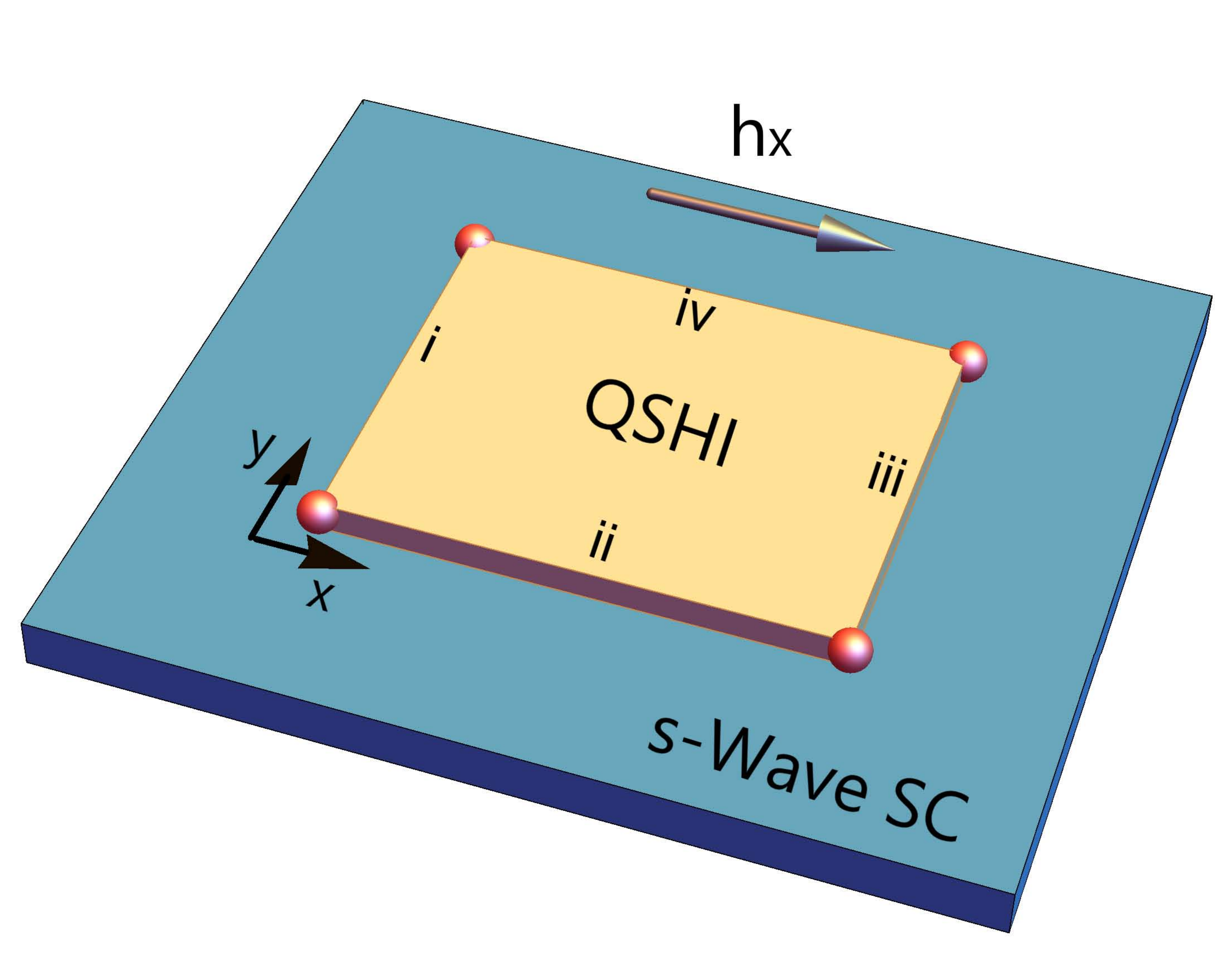}
\caption{Illustration of a heterostructure composing of a quantum spin Hall
insulator on top of an $s$-wave superconductor and subject to an in-plane
Zeeman field. The spheres at four corners represent four Majorana zero
energy modes.}
\label{Fig1}
\end{figure}

Recently, a new class of topological superconductors, dubbed as higher-order
topological superconductors, has been proposed \cite%
{Langbehn2017,ZhuX2018,Khalaf2018,Yan2018,Wang2018,
wangyu2018,Pan2018,Liu2018,Hsu2018,Yanick2018,ZhuX12018,Franca2018,Ghorashi2019, Laubscher2019}%
. In contrast to conventional topological superconductors, $r$th-order ($%
r\geq 2$) topological superconductors in $d$ dimensions host ($d-r$%
)-dimensional Majorana bound states, rather than $d-1$-dimensional gapless
Majorana excitations. For example, in 2D second-order topological
superconductors, the edge modes manifest themselves as 0D Majorana
excitations localized at the corners, instead of 1D edges, giving rise to
Majorana corner modes (MCMs). A variety of schemes have been proposed
recently to implement MCMs, such as $p$-wave superconductors under magnetic
field \cite{ZhuX2018}, 2D topological insulators in proximity to high
temperature superconductors ($d$-wave or $s_{\pm }$-wave pairing) \cite%
{Yan2018,Wang2018, Liu2018}, $\pi $-junction Rashba layers \cite{Yanick2018}
in contact with $s$-wave superconductors. However, those schemes demand
either unconventional superconducting pairings or complicated
junction/lattice structures, which are difficult to implement with current
experimental technologies.

In this Letter, we propose that MCMs can be realized with a simple and
experimentally already realized heterostructure \cite%
{Yu2014,Pribiag2015,Shi2015,Lupke2019} composing of an $s$-wave
superconductor in proximity contact with a quantum spin Hall insulator
(QSHI) and subject to an in-plane Zeeman field, as sketched in Fig.~\ref%
{Fig1}. Here we consider a simple square lattice. At each edge of the 2D
QSHI, there are two helical edge states with opposite spins and momenta,
which thus support proximity-induced $s$-wave superconducting pairing,
resulting in a quasiparticle band gap for the helical edge mode spectrum
\cite{Yu2014,Pribiag2015,Shi2015,Lupke2019}.

\begin{figure}[t]
\centering\includegraphics[width=0.48\textwidth]{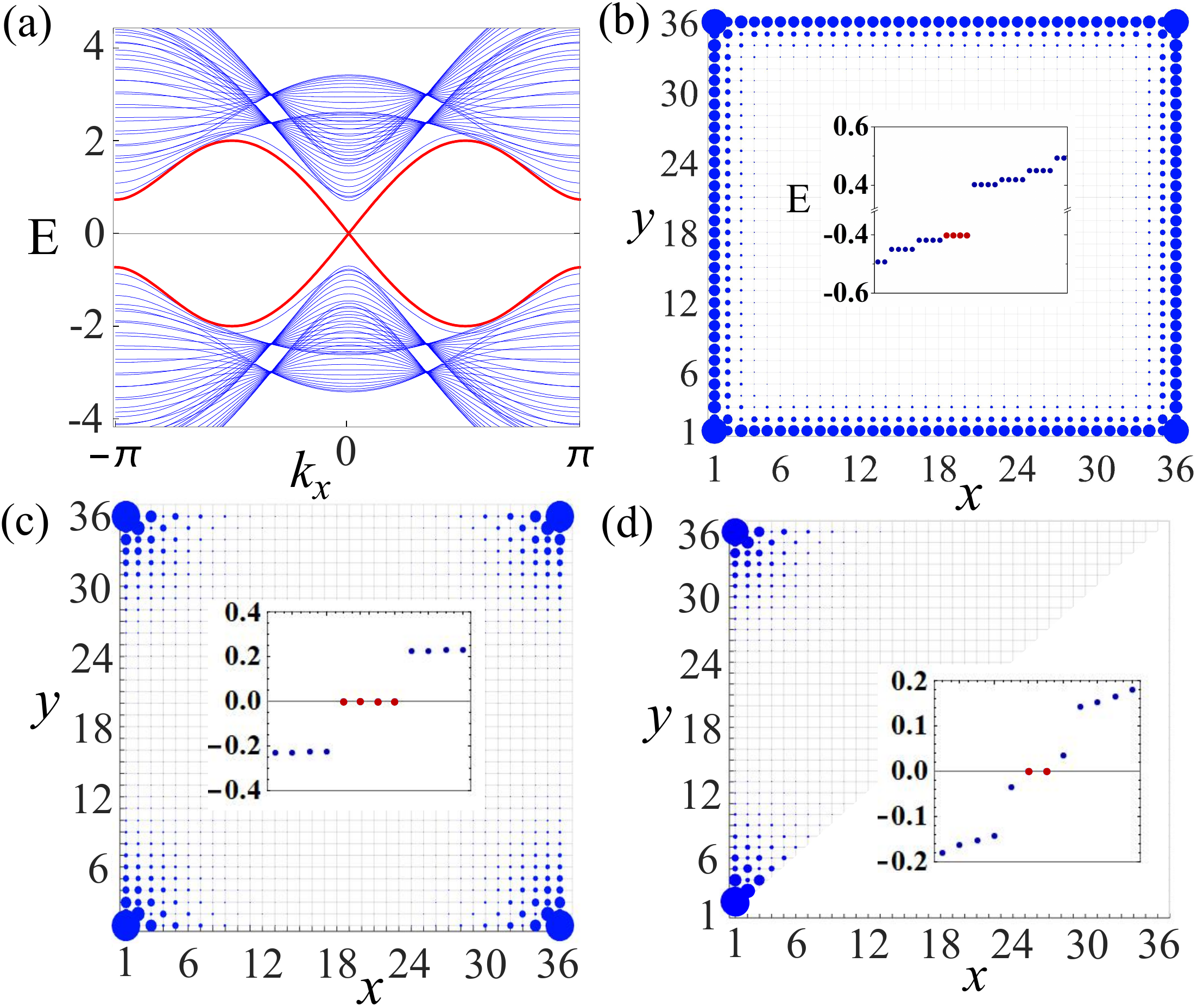}
\caption{(a) Quasiparticle bands with edge spectrum (red lines) for open
boundary conditions along the $y$ direction. The gap for edge spectrum
closes at $h_{x}=\Delta _{0}=0.4$. (b) Density distributions of the edge
bound states (red dots in the inset) for a trivial superconductor, where we
have chosen $h_{x}=0.0$, $\Delta _{0}=0.4$. (c)-(d) Density distributions of
MCMs in different geometries. The radii of the blue disks are proportional
to local density. The insets show the energy levels for $h_{x}=0.8$ and $%
\Delta _{0}=0.4$. In both Fig. \protect\ref{Fig2} and Fig. \protect\ref{Fig3}%
, $t_{x}=t_{y}=\protect\lambda _{x}=\protect\lambda _{y}=1.0$ and $\protect%
\epsilon _{0}=1.0$.}
\label{Fig2}
\end{figure}

Because of different spin-orbit coupling at adjacent edges, an in-plane
Zeeman field induces quite different effects on adjacent edges. Across a
critical Zeeman field, the quasiparticle band gap along one edge first
closes then reopens, indicating a topological phase transition, but remains
unaffected for adjacent edges. Before the phase transition, the Dirac mass
term in the low-energy effective Hamiltonian for the helical edge states has
the same sign for two adjacent edges. After the topological phase
transition, the Dirac mass term reverses its sign at the corner connecting
two edges, resulting in MCMs. In contrast, the corner Dirac mass sign change
in previous schemes originates from the sign change of the pairing order
through unconventional superconducting pairing. There is only one MCM at
each corner due to the time-reversal symmetry breaking, instead of Majorana
Kramers pairs \cite{Yan2018, Wang2018}.

Applying similar physics to three dimensions, we find that second-order
topological superconductor can be implemented in a 3D strong topological
insulator, where the interplay between $s$-wave pairing and Zeeman field
(not necessarily in-plane) gives rise to four domain walls on the edges
between two neighboring surfaces, yielding Majorana hinge modes.

{\color{blue}\emph{Physical system and model Hamiltonian.---}} Consider a
QSHI in proximity contact with an s-wave superconductor and subject to a
Zeeman field $\bm{h}$ (see Fig.~\ref{Fig1}). The four edges of a square
sample are labeled by $\mathrm{i}$, $\mathrm{ii}$, $\mathrm{iii}$, $\mathrm{%
iv}$. The physics of the heterostructure can be described by an effective
Hamiltonian \cite{Yan2018}%
\begin{eqnarray}
H(\mathbf{k}) &=&2\lambda _{x}\sin k_{x}\sigma _{x}s_{z}\tau _{z}+2\lambda
_{y}\sin k_{y}\sigma _{y}\tau _{z}  \notag \\
&&+\left( \xi _{\mathbf{k}}\sigma _{z}-\mu \right) \tau _{z}+\Delta _{0}\tau
_{x}+\bm{\mathit{h}}\cdot \bm{\mathit{s}},  \label{Ham}
\end{eqnarray}%
under the basis $\hat{C}_{\mathbf{k}}=\left( c_{\mathbf{k}},-is_{y}c_{-%
\mathbf{k}}^{\dagger }\right) ^{T}$ with $c_{\mathbf{k}}=\left( c_{\mathbf{k}%
,a,\uparrow },c_{\mathbf{k},b,\uparrow },c_{\mathbf{k},a,\downarrow },c_{%
\mathbf{k},b,\downarrow }\right) ^{T}$. Here $\lambda _{i}$ is the
spin-orbit coupling strength, $\Delta _{0}$ denotes $s$-wave superconducting
order parameter induced by proximity effect, $\xi _{\mathbf{k}}=\epsilon
_{0}-2t_{x}\cos k_{x}-2t_{y}\cos k_{y}$ with $2\epsilon _{0}$ being the
crystal-field splitting energy and $t_{i}$ the hopping strength on the
square lattice, and $\mu $ is the chemical potential. Three Pauli matrices $%
\bm{\sigma}$, $\bm{s}$ and $\bm{\tau}$ act on orbital ($a$, $b$), spin ($%
\uparrow $, $\downarrow $) and particle-hole degrees of freedom,
respectively. For simplicity of the presentation, we focus on the $\mu =0$
case, where simple analytic results for edge modes can be obtained.

In the absence of superconducting pairing and Zeeman field, the Hamiltonian (%
\ref{Ham}) is invariant under the time-reversal $\mathcal{T}=is_{y}K$ and
space-inversion $\mathcal{I}=\sigma _{z}$ operations, where $K$ is the
complex-conjugation operator. Here the band topology can be characterized by
a $Z_{2}$ topological index protected by $\mathcal{T}$ symmetry or an
equivalent $Z$ index for the spin Chern number \cite{He2016}. The system is
a QSHI in the band inverted region $\left[ \epsilon _{0}^{2}-\left(
2t_{x}+2t_{y}\right) ^{2}\right] \left[ \epsilon _{0}^{2}-\left(
2t_{x}-2t_{y}\right) ^{2}\right] <$ $0$. With the open boundary condition,
there are two helical edge states with opposite spins and momenta
propagating along each edge in the QSHI phase \cite{Hansan2010,Qix2011}.

{\color{blue}\emph{Topological phase diagram and MCMs.{--- }}}In the
presence of $\Delta _{0}$, a finite quasiparticle energy gap is opened in
the edge spectrum for two helical edge states due to the \textit{s}-wave
pairing. The in-plane Zeeman field $h_{x}$ has different effects on the
single particle edge spectra (i.e., $\Delta _{0}=0$) along the $x$ and $y$
directions: it can (cannot) open the gap along the $k_{x}$ ($k_{y}$)
direction~\cite{SM}. Such anisotropic effect of $h_{x}$ leads to very
different physics when $\Delta _{0}\neq 0$. Along the $k_{x}$ direction, the
quasiparticle band gap first closes [Fig.~\ref{Fig2}(a)] at the critical
point $h_{xc}=\Delta _{0}$ and then reopens with increasing $h_{x}$,
indicating a topological phase transition. While along the $k_{y}$
direction, the quasiparticle band gap does not close~\cite{SM}.
The difference between the edge spectra drives the heterostructure to a
second-order topological superconductor.

\begin{figure}[tbp]
\centering\includegraphics[width=0.48\textwidth]{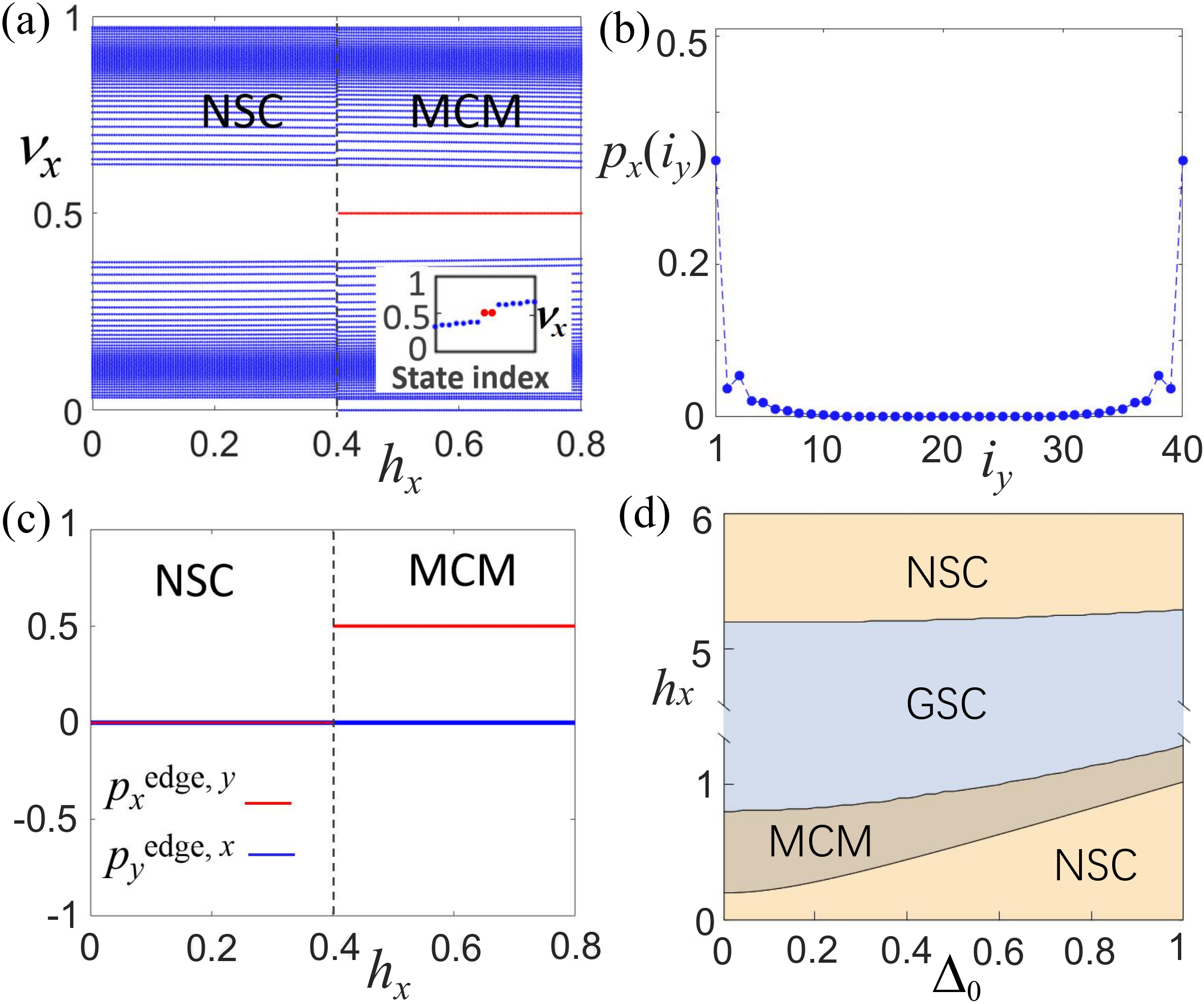}
\caption{ (a) Wannier spectra $\protect\nu _{x}$ versus $h_{x}$. The inset
showcases Wannier centers for different state indexes with $h_{x}=0.8$. (b)
Majorana polarization distribution $p_{x}\left( i_{y}\right) $ versus
lattice index $i_{y}$. (c) Majorana edge polarizations $p_{x}^{\mathrm{edge}%
,y}$ and $p_{y}^{\mathrm{edge},x}$ along $y$-normal and $x$-normal edges,
respectively. In (a)-(c), $\Delta _{0}=0.4$ and $\protect\mu =0.0$ are used.
(d) Phase diagram with $\protect\mu =0.2$. NSC denotes a trivial
superconductor, and GSC denotes a gapless superconductor. MCM and NSC phases
are separated by the phase boundary $h_{x}=\protect\sqrt{\protect\mu %
^{2}+\Delta _{0}^{2}}$.}
\label{Fig3}
\end{figure}

The emergence of MCMs after the topological phase transition is confirmed by
the numerical simulation of corresponding lattice tight-binding model in
real space, as shown in Figs. \ref{Fig2}(b), 2(c). Before the topological
phase transition [Fig. \ref{Fig2}(b)], there are no zero energy bounded
states localized at edges. After the in-plane Zeeman field exceeds the
critical point $h_{xc}$, four zero energy MCMs emerge at each corner of the
square sample [Fig. \ref{Fig2}(c)]. The emergence of MCMs is independent of
the underlying geometry of the sample. For example, in Fig. \ref{Fig2}(d), a
similar result is observed in a equilateral right triangular sample. In this
case, there are only two MCMs at two left corners due to the orientation of
the hypotenuse edge that leads to different effects of the in-plane Zeeman
field.

To examine the topological characterization of MCMs, we further calculate
the Majorana edge polarizations $p_{x}^{\mathrm{edge},y}$ and $p_{y}^{%
\mathrm{edge},x}$ using the Wilson loops on a cylindrical geometry \cite%
{Khalaf19,Benalcazar17}. Majorana edge polarization at the $y$-normal edge
is defined by $p_{x}^{\mathrm{edge},y}=\sum_{i_{y}=1}^{N_{y}/2}p_{x}\left(
i_{y}\right) $, where $N_{y}$ is the number of unit cells along $y$, and the
polarization distribution is $p_{x}\left( i_{y}\right) =\frac{1}{N_{x}}%
\sum_{j,k_{x},\beta ,n}\left\vert \left[ u_{k_{x}}^{n}\right] ^{i_{y},\beta }%
\left[ \nu _{k_{x}}^{j}\right] ^{n}\right\vert ^{2}\nu _{x}^{j}$. Here, $%
\left[ \nu _{k_{x}}^{j}\right] ^{n}$ represents the $n$th component of the $%
j $th eigenvector corresponding to the Wannier center $\nu _{x}^{j}$ of the
Wannier Hamiltonian $\mathcal{H}_{\mathcal{W}_{x}}=-i\ln \mathcal{W}_{x}$
with $\mathcal{W}_{x}$ the Wilson loop operator \cite{SM}. $\left[
u_{k_{x}}^{n}\right] ^{i_{y},\beta }$ is the $\left( i_{y},\beta \right) $%
-th component of occupied state $\left\vert u_{k_{x}}^{n}\right\rangle $
with $i_{y}$ and $\beta $ being the site index and the internal degrees of
freedom, respectively. Similarly, we can define Majorana edge polarization $%
p_{y}^{\mathrm{edge},x}$. In the MCM phase, only the Wannier spectra $\nu
_{x}$ contain two half-quantized Wannier values, as shown in Fig. \ref{Fig3}%
(a), implying that the edge polarizations occur only along the $y$-normal
edges but not the $x$-normal edges. Such an observation has been numerically
verified by distributions of localized edge polarization along $y$ [see Fig. %
\ref{Fig3}(b)], and zero edge polarization distributions along $x$. This
further leads to half quantization of $p_{x}^{\mathrm{edge},y}$ and
vanishing $p_{y}^{\mathrm{edge},x}$, as demonstrated in Fig. \ref{Fig3}(c).

We remark that the above topological characterizations show that the MCM
phase in our system falls into the class of extrinsic higher-order
topological phases distinguished by gap closings of the edge spectra \cite%
{Khalaf19} on a cylindrical geometry, instead of bulk spectra on a torus
geometry for intrinsic higher-order phases. However, the MCMs cannot be
annihilated by perturbations without closing the edge energy gap \cite{SM}.

{\color{blue}\emph{Low-energy theory on edges.{---} \ }}All above numerical
results can be explained by developing an effective low-energy theory on
edges. With both $\Delta _{0}$ and $h_{x}$, the Hamiltonian $H(\mathbf{k})$
possesses both inversion symmetry and particle-hole symmetry $\mathcal{P}H(%
\mathbf{k})\mathcal{P}^{-1}=-H(-\mathbf{k})$, but breaks the time-reversal
symmetry, where $\mathcal{P}=\tau _{x}K$. Without loss of generality, we
assume a positive in-plane Zeeman field applied along $x$ direction, i.e., $%
h_{x}>0$ and $h_{y}=h_{z}=0$. The eigenenergies of $H(\mathbf{k})$ are $%
E\left( \mathbf{k}\right) =\pm \sqrt{\left( 2\lambda _{x}\sin k_{x}\right)
^{2}+\left( \varsigma \pm h_{x}\right) ^{2}}$, where $\varsigma =\sqrt{\xi _{%
\mathbf{k}}^{2}+\left( 2\lambda _{y}\sin k_{y}\right) ^{2}+\Delta _{0}^{2}}$
and each of them is twofold degenerate. For large $h_{x}$, the system must
be a normal superconductor, which becomes gapless for moderate $h_{x}$.

When $h_{x}$ is small, the low-energy effective Hamiltonian can be obtained
through the lowest order expansion with respect to $\mathbf{k}$ at $\Gamma $
point
\begin{eqnarray}
H_{\mathrm{eff}}\left( \mathbf{k}\right) &=&\left( \epsilon
+t_{x}k_{x}^{2}+t_{y}k_{y}^{2}\right) \sigma _{z}\tau _{z}+2\lambda
_{x}k_{x}\sigma _{x}s_{z}\tau _{z}  \notag \\
&&+2\lambda _{y}k_{y}\sigma _{y}\tau _{z}+\Delta _{0}\tau _{x}+h_{x}s_{x},
\label{Ham2}
\end{eqnarray}%
where $\epsilon =\epsilon _{0}-2t_{x}-2t_{y}<0$ is assumed for topologically
nontrivial QSHI.

Assuming an open-boundary condition along the $x$ direction for edge $%
\mathrm{i}$, we can replace $k_{x}$ with $-i\partial _{x}$ and rewrite $H_{%
\mathrm{eff}}\left( \mathbf{k}\right) =H_{0}\left( -i\partial _{x}\right)
+H_{p}\left( k_{y}\right) $ with $H_{0}=\left( \epsilon -t_{x}\partial
_{x}^{2}\right) \sigma _{z}\tau _{z}-2i\lambda _{x}\partial _{x}\sigma
_{x}s_{z}\tau _{z}$, and $H_{p}=t_{y}k_{y}^{2}\sigma _{z}\tau _{z}+2\lambda
_{y}k_{y}\sigma _{y}\tau _{z}+\Delta _{0}\tau _{x}+h_{x}s_{x}$. When $\Delta
_{0}$ is small comparing to the energy gap, we can treat $H_{p}$ as a
perturbation and solve $H_{0}$ to derive the effective edge Hamiltonian for
edge $\mathrm{i}$. Assume that $\Psi _{a}$ is a zero energy solution for $%
H_{0}$ bounded at edge $\mathrm{i}$, $\sigma _{y}s_{z}\tau _{z}\Psi _{a}$ is
also the eigenstate for $H_{0}$ due to $\left\{ H_{0},\sigma
_{y}s_{z}\right\} =0$. We choose the basis vector $\zeta _{\beta }$ for $%
\Psi _{a}$ satisfying $\sigma _{y}s_{z}$ $\zeta _{\beta }=-\zeta _{\beta }$,
where $\zeta _{1}=\left\vert -,+,+\right\rangle $, $\zeta _{2}=\left\vert
+,-,+\right\rangle $, $\zeta _{3}=\left\vert -,+,-\right\rangle $, $\zeta
_{4}=\left\vert +,-,-\right\rangle $ are eigenstates of $\sigma
_{y}s_{z}\tau _{z}$. Under this basis, the effective low-energy Hamiltonian
for the edge becomes $H_{\mathrm{edge,}i}=2i\lambda _{y}s_{z}\tau
_{z}\partial _{y}+\Delta _{0}\tau _{x}$ with the topology characterized by a
$Z$ invariant \cite{Geier2018}. Similarly, we obtain the low-energy
Hamiltonian for every edge 
\begin{equation}
H_{\mathrm{edge},j}=-i\lambda _{j}s_{z}\tau _{z}\partial _{l_{j}}+\Delta
_{0}\tau _{x}+h_{j}s_{x}.  \label{ens}
\end{equation}%
Here the parameters are $\lambda _{j}=\left\{ -2\lambda _{y},2\lambda
_{x},2\lambda _{y},-2\lambda _{x}\right\} $, $l_{j}=\left\{ y,x,y,x\right\} $%
, and $h_{j}=\left\{ 0,h_{x},0,h_{x}\right\} $ for $j=\mathrm{i}$-$\mathrm{iv%
}$ edges.

From the effective edge Hamiltonian (\ref{ens}), we see that the
superconducting order induces quasiparticle gaps for all helical edge states
regardless of Zeeman fields since $\left\{ s_{z}\tau _{z},\tau _{x}\right\}
=0$. On the other hand, Eq. (\ref{ens}) indicates that the in-plane Zeeman
field $h_{x}$ only opens a gap on two parallel edges ($\mathrm{ii}$ and $%
\mathrm{iv}$), but keeps two perpendicular edges ($\mathrm{i}$ and $\mathrm{%
iii}$) untouched~\cite{SM}.

When $\Delta _{0}=0$, the low-energy edge Hamiltonian possesses two
zero-energy bound states on edge $\mathrm{i}$: $\Psi _{1}\left( x\right)
=A_{1}\left( \sin \alpha x \right) e^{-\frac{\lambda _{x}}{t_{x}}x}\left(
\zeta _{1}+\zeta _{2}\right) $ and $\Psi _{2}\left( x\right) =A_{2}\left(
\sin \alpha x \right) e^{-\frac{\lambda _{x}}{t_{x}}x}\left( \zeta
_{3}+\zeta _{4}\right) $, where $\alpha =\sqrt{-\left( \lambda
_{x}^{2}/t_{x}^{2}+h_{x}/t_{x}+\epsilon /t_{x}\right) }$ and $A_{1}$($A_{2}$%
) is the normalization constant. Similarly, there are two zero energy bound
states localized at edge $\mathrm{iii}$, which are confirmed by real space
numerical simulation~\cite{SM}.

After a unitary transformation $U=1\oplus \left( -is_{y}\right) $, the edge
Hamiltonian reads%
\begin{equation}
H_{\mathrm{edge},j}^{\prime }=-i\lambda _{j}s_{z}\partial _{l_{j}}+\Delta
_{0}s_{x}\tau _{z}+h_{j}s_{x},
\end{equation}%
on the rotated basis $\chi _{1}=\left\vert +1\right\rangle \left\vert
+1\right\rangle $, $\chi _{2}=\left\vert +1\right\rangle \left\vert
-1\right\rangle $, $\chi _{3}=\left\vert -1\right\rangle \left\vert
+1\right\rangle $, $\chi _{4}=\left\vert -1\right\rangle \left\vert
-1\right\rangle $, which are eigenstates of $s_{z}\tau _{y}$. For edge $%
\mathrm{i}$, the Hamiltonian $H_{\mathrm{edge,}i}^{\prime }$ has two
decoupled diagonal blocks with Dirac masses $\Delta _{0}+h_{x}$ and $%
h_{x}-\Delta _{0}$, respectively. While for edge $\mathrm{ii}$, Dirac masses
are the same $\Delta _{0}$ for two blocks. When $(\Delta _{0}-h_{x})\Delta
_{0}<0$ (i.e., $h_{x}>\Delta _{0}$), the Dirac masses on edges $\mathrm{i}$
and $\mathrm{ii}$ have opposite signs, leading to the emergence of a
localized mode at the intersection of two edges, which is the MCM observed
numerically in Fig. \ref{Fig2}(c). At the corner between edges $\mathrm{i}$
and $\mathrm{ii}$, the MCM can be obtained from the zero-energy wave
function
\begin{equation}
\Phi \left( x,y\right) \propto \left\{
\begin{array}{c}
e^{-\frac{\left\vert \Delta _{0}-h_{x}\right\vert }{2\lambda _{y}}\left\vert
y-y_{0}\right\vert }\left( \chi _{3}-i\chi _{4}\right) \text{ \ (edge }%
\mathrm{i}\text{)}, \\
e^{-\frac{\Delta _{0}}{2\lambda _{x}}\left\vert x-x_{0}\right\vert }\left(
\chi _{3}-i\chi _{4}\right) \text{ \ \ \ \ \ \ (edge }\mathrm{ii}\text{)},%
\end{array}%
\right.
\end{equation}%
where the corner locates at $(x_{0},y_{0})$. We see that MCMs could have
different density distributions along different directions when $\left\vert
\Delta _{0}-h_{x}\right\vert /\lambda _{y}\neq \Delta _{0}/\lambda _{x}$.

For the triangle geometry in Fig. \ref{Fig2}(d), the effect of the in-plane
Zeeman field on the hypotenuse edge can be studied by projecting it to the
direction of the Zeeman field, which shows that the Zeeman field acts
uniformly on the hypotenuse and upper edges. Consequently, there is no kink
of Dirac mass at that corner, i.e., no MCM. Generally, such an argument
applies to all geometric configurations with odd edges (e.g., a square with
a small right triangle removed at a corner), which is consistent with bulk
spectra because the particle-hole symmetry demands that zero-energy modes
must be lifted pairwise.

For a general form of the in-plane Zeeman field, MCMs emerge in the region $%
\sqrt{h_{x}^{2}+h_{y}^{2}}>\Delta _{0}$. However, an out-of-plane Zeeman
field ($h_{z}s_{z}$ term) does not induce MCMs because the helical edge
states of QSHIs remain gapless for any $h_{z}$ and $h_{z}$ affects each edge
in the same way~\cite{SM}. Finally, for a nonzero chemical potential $\mu
\neq 0$, the spectrum is more complicated~\cite{SM}, and MCMs still exist
for $h_{x}>\sqrt{\mu ^{2}+\Delta _{0}^{2}}$ with bulk spectrum being gapped.
The phase diagram with a finite $\mu $ is shown in Fig. \ref{Fig3}(d).

\begin{figure}[tbp]
\centering\includegraphics[width=0.46\textwidth]{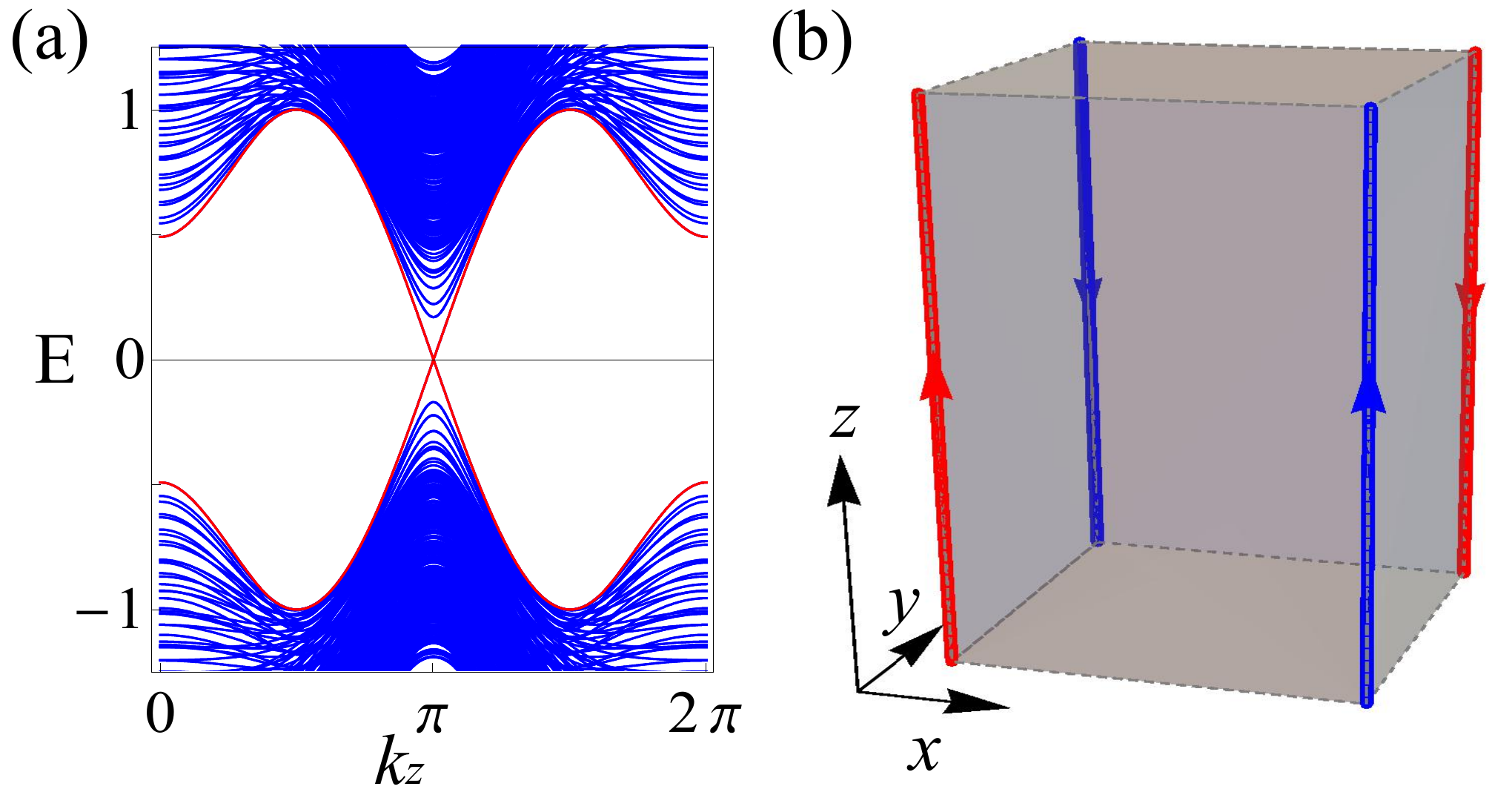}
\caption{(a) Quasiparticle spectrum along $k_{z}$ with open boundary
conditions along the $x$ and $y$ directions. (b) Majorana hinge excitations
in a 3D second-order topological superconductor. Parameters are $%
t_{x}=t_{y}=t_{z}=\protect\lambda _{x}=\protect\lambda _{y}=\protect\lambda %
_{z}=1.0$, $\Delta _{0}=0.3$, $h_{x}=0.6$, and $m_{0}=2.0$.}
\label{Fig4}
\end{figure}

{\color{blue}\emph{Majorana hinge modes in three dimensions.---}} Similar
physics also applies to three dimensions. Consider a 3D topological
insulator described by the Hamiltonian $H_{T}\left( \mathbf{k}\right) =\xi _{%
\mathbf{k}}^{\prime }\sigma _{z}s_{0}+\sum_{i}\lambda _{i}\sin k_{i}\sigma
_{x}s_{i}$ with $\xi _{\mathbf{k}}^{\prime }=m_{0}+\sum_{i}t_{i}\cos k_{i}$,
which respects both time reversal and inversion symmetries. For $%
1<\left\vert m_{0}\right\vert <3$, $H_{T}\left( \mathbf{k}\right) $
represents a 3D topological insulator that possesses surface Dirac cones
with gapped bulk spectrum protected by $\mathcal{I}$ and $\mathcal{T}$
symmetries. In the presence of an \textit{s}-wave superconducting order $%
\Delta _{0}$ and a Zeeman field,
\begin{eqnarray}
H_{3D}\left( \mathbf{k}\right)  &=&\xi _{\mathbf{k}}^{\prime }\sigma
_{z}\tau _{z}+\lambda _{x}\sin k_{x}\sigma _{x}s_{x}\tau _{z}+\lambda
_{y}\sin k_{y}\sigma _{x}s_{y}\tau _{z}  \notag \\
&&+\lambda _{z}\sin k_{z}\sigma _{x}s_{z}\tau _{z}+\Delta _{0}\tau _{x}+%
\bm{\mathit{h}}\cdot \bm{\mathit{s}}.
\end{eqnarray}%
For $\Delta _{0}\neq 0$, $\left\vert \bm{\mathit{h}}\right\vert =0$, the
surface states are gapped and the system is a trivial superconductor. When $%
h_{x}>0$, $h_{y}=h_{z}=0$, the in-plane Zeeman field $h_{x}$ breaks the time
reversal symmetry in the $x$ direction, generating a class-D superconductor.
Tuning $h_{x}>\Delta _{0}$, we observe the gapless chiral Majorana hinge
modes propagating along the $z$ direction as shown in Fig. \ref{Fig4}. Such
a 3D second-order topological superconductor can be characterized by a $Z$
invariant \cite{Geier2018}.

Figure~\ref{Fig4}(a) shows the energy spectrum with open boundary conditions
along $x$ and $y$ directions, where the chiral Majorana hinge modes (each
twofold degenerate) emerge in the bulk energy gap. The combination of the
Zeeman field and the superconductor order gives rise to four domain walls at
which the Dirac mass sign changes. Because of the inversion symmetry, the
chiral modes at diagonal hinges propagate along opposite directions, as
illustrated in Fig \ref{Fig4}(b). We remark that the requirement of in-plane
Zeeman field can be released in three dimensions and the direction of the
Zeeman field can be used to control the directionality of the hinge modes.
Specifically, when the Zeeman field lies along the $y$ ($z$) direction and $%
h_{y}>\Delta _{0}$ ($h_{z}>\Delta _{0}$), the chiral Majorana hinge modes
propagate along the $x$ ($y$) direction with periodic boundary conditions.

{\color{blue}\emph{Discussion and conclusion.---}}InAs/GaSb quantum wells
are 2D $Z_{2}$ QSHIs with large bulk insulating gaps up to $\sim 50$ meV,
and significant experimental progress has been made \cite%
{Du2017a,Nichele2017,Du2017b,Du2015} recently to observe their helical edge
states. Superconducting proximity effects in InAs/GaSb quantum wells were
also observed in experiments \cite{Yu2014,Pribiag2015,Shi2015}. In
particular, edge-mode superconductivity due to proximity contact with an $s$%
-wave superconductor has been detected through transport measurement \cite%
{Pribiag2015}, and giant supercurrent states have been observed \cite%
{Shi2015}. The in-plane Zeeman field could be realized using an in-plane
magnetic field due to the relatively large $g$ factor for InAs/GaSb quantum
wells \cite{Mu2016}. By engineering a suitable quantum device, zero-bias
peaks for MCMs should be observable in transport or STM types of experiments.

Another potential material is the monolayer WTe$_{2}$ \cite{Qian2014} that
has been confirmed as a QSHI in recent experiments \cite{Wu2018,Shi2019}.
When proximate to superconductors, a proximity-induced superconducting gap
of the order of $\sim 0.7$~meV \cite{Lupke2019} emerges. To achieve a
comparable spin Zeeman splitting, an in-plane magnetic field $H\sim 0.3$--$3$%
~T is required, given the Land\'{e} \emph{g} factor ranges from 4.5 \cite%
{Wu2018,Aivazian2015} to larger than 44 \cite{Bi2018} in WTe$_{2}$,
depending on the direction of the applied fields. Based on the
aforementioned parameters, \emph{s}-wave superconductor NbN can be used for
the device fabrication, given its both high transition temperature, $%
T_{c}\sim 12$~K, and high critical field, for example, $H_{c}>12$~T at 0.5~K
\cite{Xiaoyan2019}. For Majorana hinge modes in three dimensions, effective
Zeeman fields could be induced by doping magnetic impurities into 3D
topological insulators.


In conclusion, we have shown that a heterostructure composing of QSHI/$s$%
-wave superconductor can become a second-order topological superconductor
with MCMs in the presence of an in-plane Zeeman field. Because neither
exotic superconducting pairings nor complex junction structures are
required, our scheme provides a simple and realistic platform for the
experimental study of the non-Abelian Majorana corner and hinge modes.

\begin{acknowledgments}
This work is supported by Air Force Office of Scientific Research
(FA9550-16-1-0387), National Science Foundation (PHY-1806227), and Army
Research Office (W911NF-17-1-0128). The work by C.Z. was performed in part
at the Aspen Center for Physics, which is supported by National Science
Foundation Grant No. PHY-1607611. This work is also supported in part by
NSFC under the Grant No. 11504285, and the Scientific Research Program
Funded by the Natural Science Basic Research Plan in Shaanxi Province of
China (Program No. 2018JQ1058), the Scientific Research Program Funded by
Shaanxi Provincial Education Department under the Grant No. 18JK0397, and
the scholarship from China Scholarship Council (CSC) (Program No.
201708615072).
\end{acknowledgments}

\newpage \clearpage
\onecolumngrid
\appendix

\section{Supplemental Materials for \textquotedblleft In-Plane Zeeman Field-Induced Majorana Corner and Hinge Modes in an $s$-Wave Superconductor
Heterostructure\textquotedblright}

\begin{figure}[h]
\centering\includegraphics[width=0.98\textwidth]{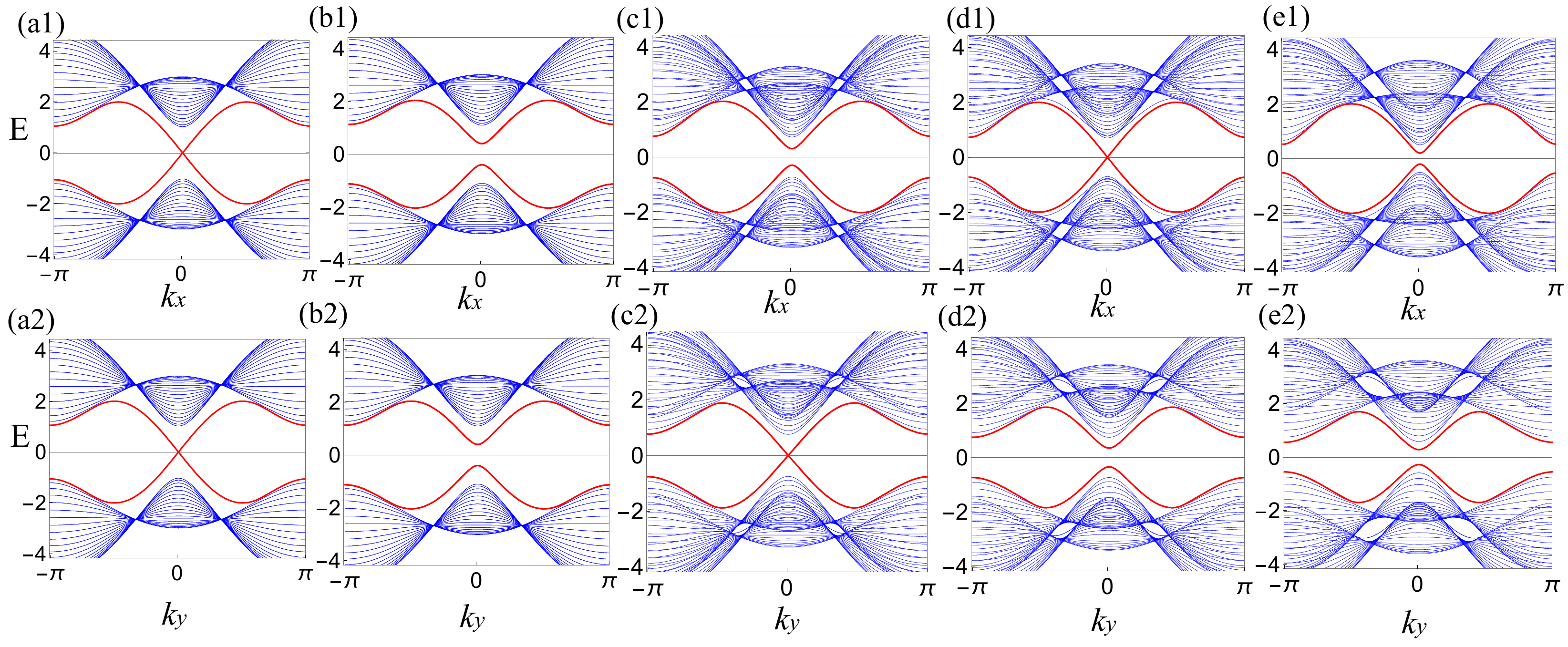}
\caption{Quasiparticle bands with edge spectrum (red lines) for open
boundary conditions along $y$ or $x$ directions. (a1) and (a2) $h_{x}=\Delta
_{0}=0.0$; (b1) and (b2) $h_{x}=0.0,\Delta _{0}=0.4$; (c1) and (c2) $%
h_{x}=0.3,\Delta _{0}=0.0$; (d1) and (d2) $h_{x}=0.4,\Delta _{0}=0.4$; (e1)
and (e2) $h_{x}=0.6,\Delta _{0}=0.4$. Other parameters: $t_{x}=t_{y}=\protect%
\lambda _{x}=\protect\lambda _{y}=1.0$, and $\protect\epsilon _{0}=1.0$.}
\label{FigS0}
\end{figure}

{\emph{Low energy edge states.--- }}The QSHI possesses helical edge states,
as shown in Figs. \ref{FigS0} (a$1$) and (a$2$). With a finite
superconducting order parameter $\Delta _{0}$, two helical edge states are
gapped due to the \textit{s}-wave pairing, as illustrated in Fig. \ref{FigS0}
(b). The in-plane Zeeman field $h_{x}$ exhibits different effects on the
single particle edge spectra along $x$ and $y$ directions (i.e., when $%
\Delta _{0}=0$): the in-plane Zeeman field can (cannot) open the edge gap
along $k_{x}$ ($k_{y}$) direction (\ref{FigS0} (c)). Such anisotropic effect
of $h_{x}$ leads to quite different physics with a finite $\Delta _{0}$.
Along the $k_{x}$ direction, the quasiparticle band gap first closes (Fig.~%
\ref{FigS0}(d$1$)) at the critical point $h_{xc}=\Delta _{0}$ and then
reopens (Fig.~\ref{FigS0}(e$1$)) with increasing $h_{x}$, which indicates
that a topological phase transition occurs. While along the $k_{y}$
direction, the quasiparticle band remains gapped (Fig.~\ref{FigS0}(d$2$, e$2$%
)).
The difference between the edge spectra along $k_{x}$ and $k_{y}$ directions
drives the heterostructure to a second-order topological superconductor,
where four zero energy MCMs emerge at each corner of the square sample.

\begin{figure}[t]
\centering\includegraphics[width=0.72\textwidth]{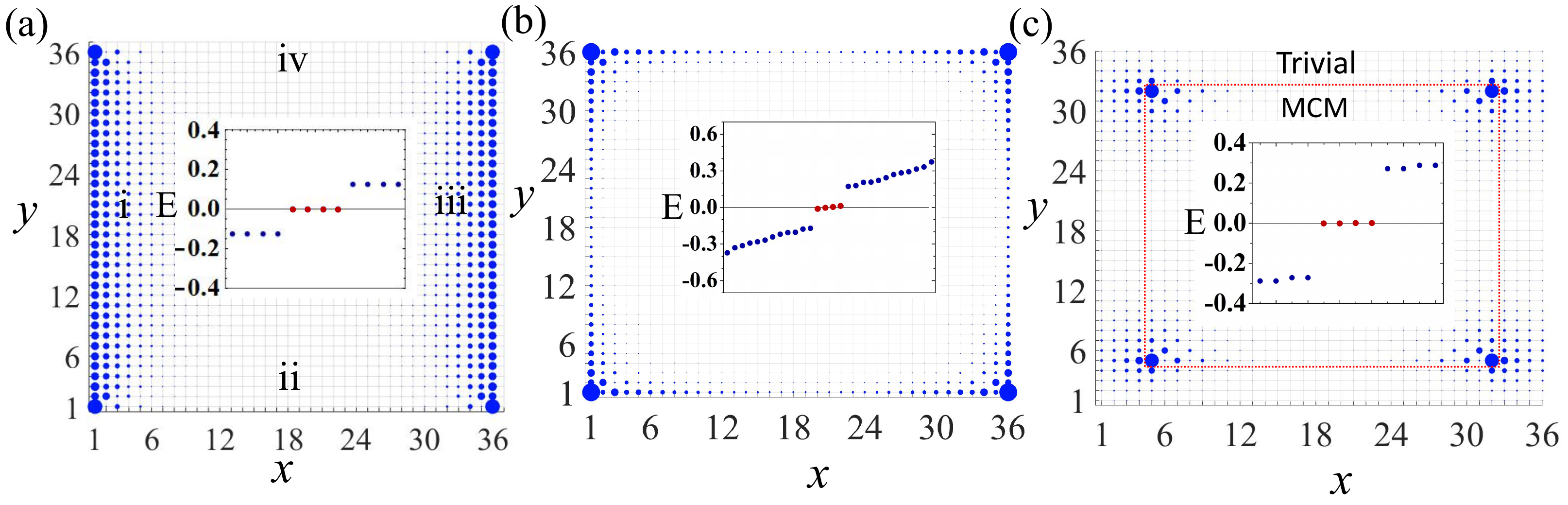}
\caption{(a) Density distributions of the zero-energy bound states when $%
\Delta _{0}=\protect\mu =0.0$, $h_{x}=0.6$. (b) Density distributions of
MCMs when $\Delta _{0}=0.2$, $\protect\mu =0.2$, $h_{x}=0.4$. The radii of
the blue disks are proportional to local density. Parameters are $%
t_{x}=t_{y}=\protect\lambda _{x}=\protect\lambda _{y}=1.0$ and $\protect%
\epsilon _{0}=1.0$. (c) Density distributions of Majorana corner modes, with
$\protect\epsilon _{0}=5.0$ outside the red dashed square (trivial phase)
and $\protect\epsilon _{0}=1.0$ inside (MCM phase). Other parameters are $%
\Delta _{0}=0.4$, $h_{x}=0.8$, $\protect\mu =0.0$.}
\label{FigS1}
\end{figure}
\begin{figure*}[t]
\centering\includegraphics[width=0.98\textwidth]{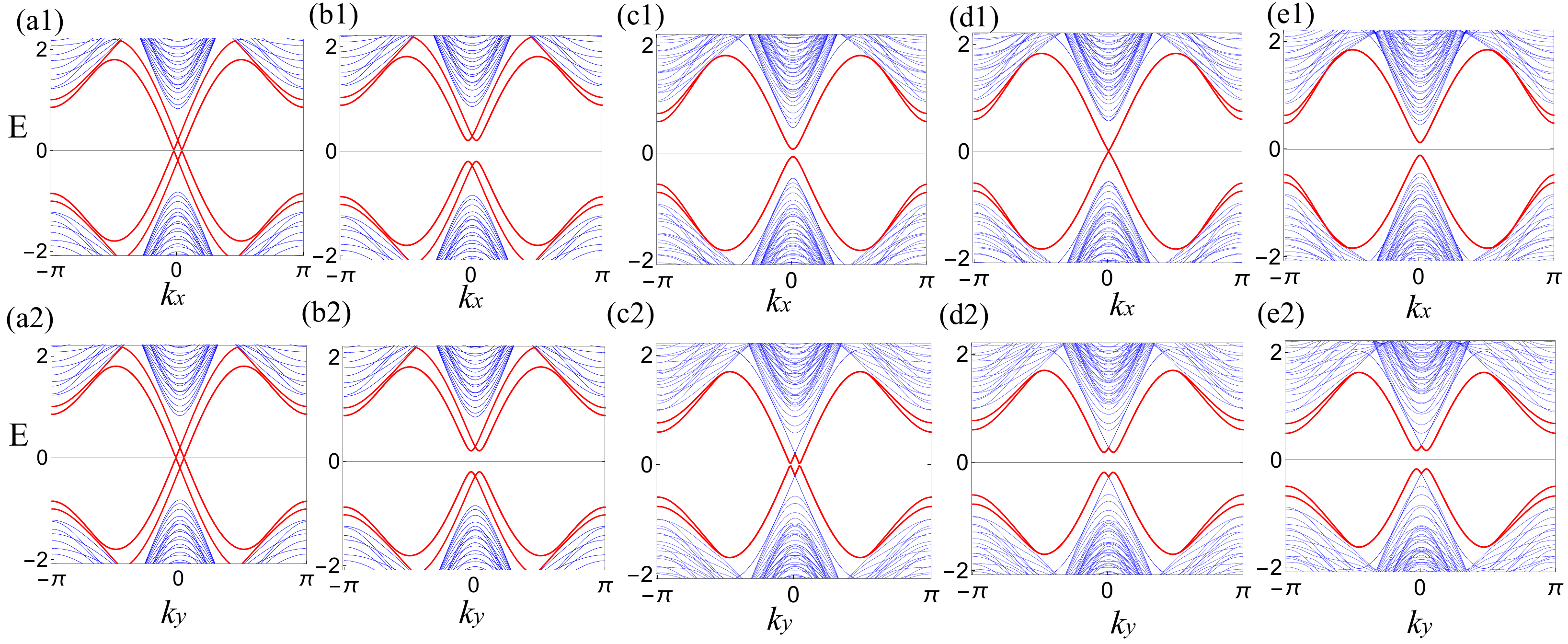}
\caption{Quasiparticle bands with edge spectrum for open boundary conditions
along $y$ or $x$ directions with a finite chemical potential $\protect\mu %
=0.2$. (a1) and (a2) $h_{x}=\Delta _{0}=0.0$; (b1) and (b2) $h_{x}=0.0$, $%
\Delta _{0}=0.2$; (c1) and (c2) $h_{x}=0.27$, $\Delta _{0}=0.0$; (d1) and
(d2) $h_{x}=0.283$, $\Delta _{0}=0.2$; (e1) and (e2) $h_{x}=0.4$, $\Delta
_{0}=0.2$. Other parameters: $t_{x}=t_{y}=\protect\lambda _{x}=\protect%
\lambda _{y}=1.0$, $\protect\epsilon _{0}=1.0$.}
\label{FigS2}
\end{figure*}

When $\Delta _{0}=\mu =0$ and in-plane Zeeman field $h_{x}\neq 0$, the low
energy states on edges ii and iv are gapped. However, zero-energy bound
states still exist on edges i and iii of the sample, as illustrated in Fig. %
\ref{FigS1} (a).

When the chemical potential $\mu \neq 0$, the superconducting pairing still
leads to a gap on the helical edge states along both $x$ and $y$ directions,
as shown in Fig. \ref{FigS2} (b). The in-plane Zeeman field induces a gap
for edge states along the $x$ direction, as illustrated in Fig. \ref{FigS2}
(c1), but it cannot open a gap for edge states along the $y$ direction, as
shown in Fig. \ref{FigS2} (c2). In the presence of both superconducting
order and in-plane Zeeman field, the low-energy edge states are in a gapped
trivial phase for $h_{x}<\sqrt{\mu ^{2}+\Delta _{0}^{2}}$ with no MCMs. At $%
h_{x}=$ $\sqrt{\mu ^{2}+\Delta _{0}^{2}}$, a topological phase transition
occurs on edge states along the $x$ direction, as illustrated in Fig. \ref%
{FigS2} (d1), but the edge modes along the $y$ direction are still gapped.
When $h_{x}>$ $\sqrt{\mu ^{2}+\Delta _{0}^{2}}$, the edge energy gap along
the $x$ direction reopens, as shown in Fig. \ref{FigS2}(e1). In this
process, the low-energy edge states along the $y$-direction remain in a
trivial gapped phase, as sketched in Fig. \ref{FigS2} (e2). In the region $%
h_{x}>$ $\sqrt{\mu ^{2}+\Delta _{0}^{2}}$, the system becomes a second order
topological superconductor, where MCMs emerge as shown in Fig. \ref{FigS1}
(b). Note that due to finite size effect, there still exist small energy
splittings for MCMs.

We emphasize that the MCMs cannot be annihilated by some perturbations that
only drive the edge region to a trivial phase. Instead, the MCMs only shift
towards inside. To confirm this, we first prepare a square sample with
Majorana corner modes, then tune parameter to make the region outside the
dashed red square in Fig.~\ref{FigS1} (c) to be in a trivial phase. By
numerical calculations, we find corner modes still exist and only recede
from previous corners. Such boundary-obstructed topological phases are
extrinsic with topological obstructions lying on the edges and the change of
topology is not associated with bulk-gap closing.

\emph{The effect of out-of-plane Zeeman field on edge states}{\emph{.---}}In
the main text, we state that out-of-plane Zeeman field does not induce a
topological phase transition to a higher-order phase, we take a closer look
at this problem here.

The ribbon Hamiltonians with periodic boundary condition along $x$/$y$
direction are respectively given by%
\begin{equation}
H=\sum_{k_{x}}H_{1}\left( k_{x}\right) +H_{2}\left( k_{x}\right)
+H_{3}\left( k_{x}\right) +H_{4}\left( k_{x}\right) ,
\end{equation}%
where%
\begin{eqnarray*}
H_{1}\left( k_{x}\right) &=&\sum_{i_{y},\sigma }\left[ \left( \epsilon
_{0}-2t_{x}\cos k_{x}\right) \left( c_{k_{x},i_{y},a,\sigma }^{\dagger
}c_{k_{x},i_{y},a,\sigma }-c_{k_{x},i_{y},b,\sigma }^{\dagger
}c_{k_{x},i_{y},b,\sigma }\right) -t_{y}\left( c_{k_{x},i_{y},a,\sigma
}^{\dagger }c_{k_{x},i_{y}\pm 1,a,\sigma }-c_{k_{x},i_{y},b,\sigma
}^{\dagger }c_{k_{x},i_{y}\pm 1,b,\sigma }\right) \right] , \\
H_{2}\left( k_{x}\right) &=&2\lambda _{x}\sum_{i_{y}}\sin k_{x}\left(
c_{k_{x},i_{y},a,\uparrow }^{\dagger }c_{k_{x},i_{y},b,\uparrow
}-c_{k_{x},i_{y},a,\downarrow }^{\dagger }c_{k_{x},i_{y},b,\downarrow
}-c_{k_{x},i_{y},b,\uparrow }^{\dagger }c_{k_{x},i_{y},a,\uparrow
}+c_{k_{x},i_{y},b,\downarrow }^{\dagger }c_{k_{x},i_{y},a,\downarrow
}\right) ,
\end{eqnarray*}%
\begin{eqnarray*}
H_{3}\left( k_{x}\right) &=&\lambda _{y}\sum_{i_{y}}\left(
-c_{k_{x},i_{y},a,\uparrow }^{\dagger }c_{k_{x},i_{y}+1,b,\uparrow
}+c_{k_{x},i_{y},a,\uparrow }^{\dagger }c_{k_{x},i_{y}-1,b,\uparrow
}-c_{k_{x},i_{y},a,\downarrow }^{\dagger }c_{k_{x},i_{y}+1,b,\downarrow
}+c_{k_{x},i_{y},a,\downarrow }^{\dagger }c_{k_{x},i_{y}-1,b,\downarrow
}\right. \\
&&\left. +c_{k_{x},i_{y},b,\uparrow }^{\dagger }c_{k_{x},i_{y}+1,a,\uparrow
}-c_{k_{x},i_{y},b,\uparrow }^{\dagger }c_{k_{x},i_{y}-1,a,\uparrow
}+c_{k_{x},i_{y},b,\downarrow }^{\dagger }c_{k_{x},i_{y}+1,a,\downarrow
}-c_{k_{x},i_{y},b,\downarrow }^{\dagger }c_{k_{x},i_{y}-1,a,\downarrow
}\right) , \\
H_{4}\left( k_{x}\right) &=&h_{z}\sum_{i_{y}}\left(
c_{k_{x},i_{y},a,\uparrow }^{\dagger }c_{k_{x},i_{y},a,\uparrow
}-c_{k_{x},i_{y},a,\downarrow }^{\dagger }c_{k_{x},i_{y},a,\downarrow
}+c_{k_{x},i_{y},b,\uparrow }^{\dagger }c_{k_{x},i_{y},b,\uparrow
}-c_{k_{x},i_{y},b,\downarrow }^{\dagger }c_{k_{x},i_{y},b,\downarrow
}\right) ,
\end{eqnarray*}%
and%
\begin{equation}
H=\sum_{k_{y}}H_{1}\left( k_{y}\right) +H_{2}\left( k_{y}\right)
+H_{3}\left( k_{y}\right) +H_{4}\left( k_{y}\right) ,
\end{equation}%
where%
\begin{eqnarray*}
H_{1}\left( k_{y}\right) &=&\sum_{i_{x},\sigma }\left[ \left( \epsilon
_{0}-2t_{y}\cos k_{y}\right) \left( c_{k_{y},i_{x},a,\sigma }^{\dagger
}c_{k_{y},i_{x},a,\sigma }-c_{k_{y},i_{x},b,\sigma }^{\dagger
}c_{k_{x},i_{y},b,\sigma }\right) -t_{x}\left( c_{k_{y},i_{x},a,\sigma
}^{\dagger }c_{k_{y},i_{x}\pm 1,a,\sigma }-c_{k_{y},i_{x},b,\sigma
}^{\dagger }c_{k_{y},i_{x}\pm 1,b,\sigma }\right) \right] , \\
H_{2}\left( k_{y}\right) &=&-i\lambda _{x}\sum_{i_{x}}\left(
c_{k_{y},i_{x},a,\uparrow }^{\dagger }c_{k_{x},i_{x}+1,b,\uparrow
}-c_{k_{y},i_{x},a,\uparrow }^{\dagger }c_{k_{y},i_{x}-1,b,\uparrow
}-c_{k_{y},i_{x},a,\downarrow }^{\dagger }c_{k_{y},i_{x}+1,b,\downarrow
}+c_{k_{y},i_{x},a,\downarrow }^{\dagger }c_{k_{y},i_{x}-1,b,\downarrow
}\right. \\
&&\left. -c_{k_{y},i_{x},b,\uparrow }^{\dagger }c_{k_{y},i_{x}-1,a,\uparrow
}+c_{k_{y},i_{x},b,\uparrow }^{\dagger }c_{k_{y},i_{x}+1,a,\uparrow
}+c_{k_{y},i_{x},b,\downarrow }^{\dagger }c_{k_{y},i_{x}-1,a,\downarrow
}-c_{k_{y},i_{x},b,\downarrow }^{\dagger }c_{k_{y},i_{x}+1,a,\downarrow
}\right) , \\
H_{3}\left( k_{y}\right) &=&-2i\lambda _{y}\sin k_{y}\sum_{i_{x}}\left(
c_{k_{y},i_{x},a,\uparrow }^{\dagger }c_{k_{y},i_{x},b,\uparrow
}+c_{k_{y},i_{x},a,\downarrow }^{\dagger }c_{k_{y},i_{x},b,\downarrow
}-c_{k_{y},i_{x},b,\uparrow }^{\dagger }c_{k_{y},i_{x},a,\uparrow
}-c_{k_{y},i_{x},b,\downarrow }^{\dagger }c_{k_{y},i_{x},a,\downarrow
}\right) , \\
H_{4}\left( k_{y}\right) &=&h_{z}\sum_{i_{x}}\left(
c_{k_{y},i_{x},a,\uparrow }^{\dagger }c_{k_{y},i_{x},a,\uparrow
}-c_{k_{y},i_{x},a,\downarrow }^{\dagger }c_{k_{y},i_{x},a,\downarrow
}+c_{k_{y},i_{x},b,\uparrow }^{\dagger }c_{k_{y},i_{x},b,\uparrow
}-c_{k_{y},i_{x},b,\downarrow }^{\dagger }c_{k_{y},i_{x},b,\downarrow
}\right) .
\end{eqnarray*}

\begin{figure*}[t]
\centering\includegraphics[width=0.98\textwidth]{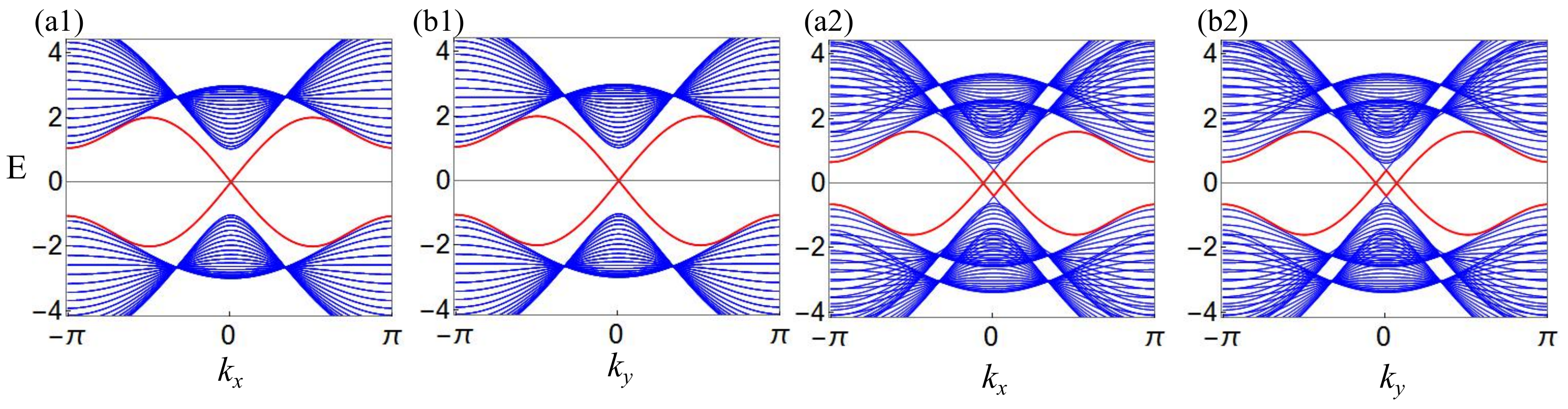}
\caption{Quasiparticle spectrums with edge states (red lines) for
open-boundary conditions along $y$ or $x$ directions. We set $h_{z}=0.0$ in
panels (a1, b1) and $h_{z}=0.4$ in panels (a2, b2). Other parameters: $%
t_{x}=t_{y}=\protect\lambda _{x}=\protect\lambda _{y}=1.0$, $\protect%
\epsilon _{0}=1.0$.}
\label{FigS3}
\end{figure*}

Through numeric calculations, we obtain energy levels for the ribbon
Hamiltonians as shown in Fig. \ref{FigS3}. Comparing with Fig. \ref{FigS3}
(a)-(b), (c)-(d) show that the $h_{z}s_{z}$ term shifts the quasiparticle
bands up and down for $\pm h_{z}$, but cannot open a gap on the helical edge
states. Therefore the out-of-plane Zeeman field $h_{z}$ acts the same on
edge states of adjacent edges. In addition, the role of $h_{z}s_{z}$ term
could also be understood from the low-energy theory of the edge states,
which can be described by the following Hamiltonian%
\begin{equation}
H_{Edge,j}=-i\lambda _{j}s_{z}\partial _{l_{j}}+h_{z}s_{z},
\end{equation}%
where $\lambda _{j}=\left\{ -2\lambda _{y},2\lambda _{x},2\lambda
_{y},-2\lambda _{x}\right\} $ and $l_{j}=\left\{ y,x,y,x\right\} $. This
edge Hamiltonian indicates that the energies of edge states are shifted by $%
h_{z}s_{z}$, which is consistent with the numeric results illustrated in
Fig. \ref{FigS3} (c) and (d). Hence, the out of plane Zeeman field merely
shifts the helical edge states and cannot open any gaps.

Finally, the energy spectra for the system with periodic boundary conditions
are written as%
\begin{equation}
E\left( k\right) =\pm \sqrt{\xi _{k}^{2}+\left( 2\lambda _{x}\sin
k_{x}\right) ^{2}+\left( 2\lambda _{y}\sin k_{y}\right) ^{2}}\pm h_{z}.
\end{equation}%
Each of them is two-fold degenerate. This energy spectra also suggest that
the $h_{z}s_{z}$ term acts the same on both helical edge states on adjacent
edges.

\begin{figure}[t]
\centering\includegraphics[width=0.98\textwidth]{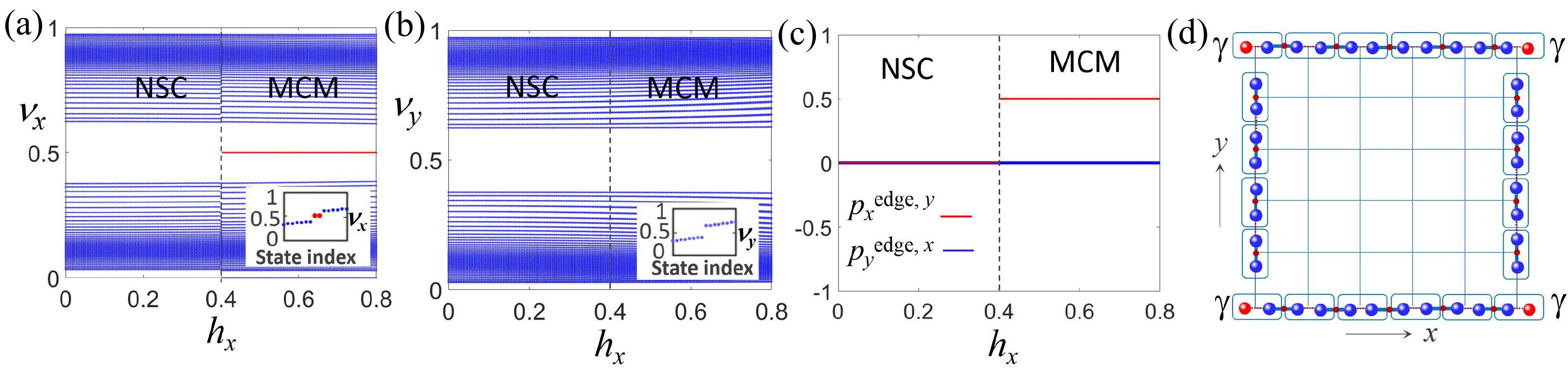}
\caption{(a) Wannier spectra $\protect\nu _{x}$ versus in-plane Zeeman field
$h_{x}$. The inset shows Wannier centers $\protect\nu _{x}$ for different
state indexes with $h_{x}=0.8$. (b) Similar to panel (a) but plotted with
Wannier spectra $\protect\nu _{y}$. (c) Majorana edge polarizations versus $%
h_{x}$. The red line denotes Majorana edge polarization $p_{x}^{\mathrm{edge}%
,y}$ at $y$-normal edge, and the blue line denotes Majorana edge
polarization $p_{y}^{\mathrm{edge},x}$ at $x$-normal edge. (d) Illustration
of four isolated Majorana corner modes $\protect\gamma $. Two Majorana
fermions in the box are combined into a Dirac fermion on a lattice site. The
red disks indicate Majorana Wannier centers. System parameters are $%
t_{x}=t_{y}=\protect\lambda _{x}=\protect\lambda _{y}=1.0$, $\protect%
\epsilon _{0}=1.0$, $\Delta _{0}=0.4$, and $\protect\mu =0.0$.}
\label{FigS4}
\end{figure}

\emph{Majorana edge polarizations}{\emph{.---}}To show the emergence of
Majorana corner modes, we calculate Majorana edge polarizations $p_{x}^{%
\mathrm{edge},y}$ and $p_{y}^{\mathrm{edge},x}$ under corresponding open
boundary conditions (open along $y$ and $x$ respectively) by the Wilson
loop. In a torus geometry with periodic boundary condition along $x$ but
open boundary condition along $y$, we consider the Wilson loop $\mathcal{W}%
_{x}=F_{x,k_{x}+\left( N_{x}-1\right) \Delta k_{x}}...F_{x,k_{x}+\Delta
k_{x}}F_{x,k_{x}}$. Here, $\left[ F_{x,k_{x}}\right] ^{mn}=\left\langle
u_{k_{x}+\Delta k_{x}}^{m}|u_{k_{x}}^{n}\right\rangle $, where $\Delta
k_{x}=2\pi /N_{x}$ with $N_{x}$ the number of unit cell along the $x$
direction, and $\left\vert u_{k_{x}}^{n}\right\rangle $ is the occupied
eigenstate of the Hamiltonian with $n$ the band index. The Wannier
Hamiltonian is defined as $\mathcal{H}_{\mathcal{W}_{x}}=-i\ln \mathcal{W}%
_{x}$, whose eigenvalues $2\pi \nu _{x}$ correspond to the Wannier spectrum,
where $\nu _{x}\equiv$mod$\left( \nu _{x},1\right) $ is the Wannier center.
Following similar procedure the Wannier spectrum $\nu _{y}$ could be also
derived. Fig. \ref{FigS4}(a) and (b) illustrate Wannier spectra $\nu _{x}$
and $\nu _{y}$ with the increasing Zeeman field, showing that in trivial
superconductor (NSC) phase, the Wannier spectra $\nu _{x,y}$ are gapped
around $1/2$, while in higher-order topological superconductor (MCM) phase,
the spectra $\nu _{x}$ exhibit two isolated eigenvalues $\nu _{x}=1/2$ (see
the inset of Fig. \ref{FigS4}(a)) but $\nu _{y}$ exhibit no isolated
eigenvalues (see the inset of Fig. \ref{FigS4}(b)). Since these isolated
eigenvalues would disappear under periodic boundary conditions along both
directions, this suggests that they originate from boundary states. In
addition, in MCM phase only $\nu _{x}=1/2$ exists but not for $\nu _{y}$
which implies that Majorana edge polarizations only occurs at the $y$-normal
edge but not at $x$-normal edge. In the following, we calculate Majorana
edge polarizations.

Majorana edge polarization at $y$-normal edge is defined by $p_{x}^{\mathrm{%
edge},y}=\sum_{i_{y}=1}^{N_{y}/2}p_{x}\left( i_{y}\right) $, where $N_{y}$
is the number of unit cells along $y$, and the polarization distribution is%
\begin{equation}
p_{x}\left( i_{y}\right) =\frac{1}{N_{x}}\sum_{j,k_{x},\beta ,n}\left\vert %
\left[ u_{k_{x}}^{n}\right] ^{i_{y},\beta }\left[ \nu _{k_{x}}^{j}\right]
^{n}\right\vert ^{2}\nu _{x}^{j}.
\end{equation}%
Here, $\left[ \nu _{k_{x}}^{j}\right] ^{n}$ represents the $n$-th component
of the $j$-th eigenvector corresponding to the Wannier center $\nu _{x}^{j}$
of the Wannier Hamiltonian $\mathcal{H}_{\mathcal{W}_{x}}$, $\left[
u_{k_{x}}^{n}\right] ^{i_{y},\beta }$ is the $\left( i_{y},\beta \right) $%
-th component of occupied state $\left\vert u_{k_{x}}^{n}\right\rangle $
with $i_{y}$ and $\beta $ being the site index and the internal degrees of
freedom, respectively. Majorana edge polarization $p_{y}^{\mathrm{edge},x}$
takes similar formulation as $p_{x}^{\mathrm{edge},y}$. Fig. \ref{FigS4}(c)
illustrates the numeric results of Majorana edge polarizations versus $h_{x}$%
, showing that in the NSC phase ($h_{x}<\Delta _{0}$), $p_{x}^{\mathrm{edge}%
,y}=p_{y}^{\mathrm{edge},x}=0$, while in the MCM phase ($h_{x}>\Delta _{0}$%
), $p_{x}^{\mathrm{edge},y}$ is quantized to $1/2$ and $p_{y}^{\mathrm{edge}%
,x}$ is quantized to $0$. This implies in MCM phase, the edge states along $%
x $ are in the topological phase in analogy to the Kitaev chain, giving rise
to four isolated Majorana corner modes, while that along $y$ is in a trivial
phase, where two Majorana fermions are combined into a complex fermion at
each lattice site, as sketched in Fig. \ref{FigS4}(d).

\end{document}